\input epsf.sty
\documentclass{kapproc} % Computer Modern font calls
\kluwerbib
\begin{document}
\articletitle{Host Galaxies and the Spectral Variability of Quasars}

\author{Fausto Vagnetti}
\affil{Dipartimento di Fisica, Universit\`a di Roma ``Tor Vergata''\\
Via della Ricerca Scientifica 1, I-00133 Roma, Italy}
\email{fausto.vagnetti@roma2.infn.it}

\author{Dario Tr\`evese}
\affil{Dipartimento di Fisica, Universit\`a di Roma ``La Sapienza''\\
Piazzale A. Moro 2, I-00185 Roma, Italy}
\email{dario.trevese@roma1.infn.it}

\section{Introduction}

Variability of the spectral energy distribution (SED) of active galactic nuclei (AGNs)
can provide clues for understanding both the main emission processes and the
origin of their variations.
So far, multi-wavelength monitoring has been possible  for a limited number of
bright objects, thanks to large international cooperations (see Ulrich, Maraschi, \& Urry 1997, and refs. therein).
The ``ensemble'' analysis of the light curves
was obtained in the past only from  single-band observations of statistical samples of QSOs (Giallongo, Tr\`evese, \& Vagnetti 1991;
Hook et al. 1994; Cristiani et al. 1996; Di Clemente et al. 1996). 
Cutri et al. (1985),  Kinney et al. (1991), and  Edelson et al. (1990),
found a  hardening of the spectrum in the bright phase in small QSO samples. 
A positive correlation of the spectral index $\alpha$ ($f_{\nu} \propto \nu^{\alpha}$)
with brightness variations $\Delta \log f_{\nu}$ has been found by
Tr\`evese et al. (2000).
The recent publication, by Giveon et al. 1999, of two-band monitoring of 42  PG QSOs,
allows a statistical study of the spectral variability of the entire set of light curves (Tr\`evese \& Vagnetti 2001).
They find a positive correlation of the $B-R$ color with the brightness variations 
$\delta B$.
These results  confirm the interpretation of the variability-redshift (v-$z$) correlation
suggested by Giallongo, Tr\`evese, \& Vagnetti (1991) and Di Clemente et al. (1996).
An independent analysis of the same light curves, performed by Cid Fernades et al. (2000) implies the existence of
an underlying spectral component, redder than the variable one, 
which could be identified either with the host galaxy,
considered by Romano \& Peterson (1998), or with a non-flaring part of the QSO spectrum. 
We evaluate the effect of the host galaxy constant SED 
on the variable QSO+host SED, through numerical simulation 
based on templates of the QSO and host SEDs
provided by Elvis et al. 1994. We also compare the results with the 
spectral variations deduced from Giveon et al. (1999),
concluding that the nuclear spectral variability is dominant.

\section{Discussion and Conclusions}

From an atlas of 47 normal QSO continuum spectra, 
Elvis et al. (1994) derive the average
QSO template spectrum corrected for the 
contribution of a template host galaxy spectrum.
On the basis of these templates we produce a synthetic (QSO+host) SED.
We characterize the relative contribution of the nucleus and the host by the parameter: $\eta \equiv \log({L_H^Q}/{L_H^g})$,
where ${L_H^Q}$ and ${L_H^g}$ are the total $H$ band luminosities of the  
QSO and the host galaxy respectively. 
To test (disprove) the hypothesis that the spectral energy distribution
of the active nucleus maintains its shape while changing brightness,
we represent variability by small changes of $\Delta\eta(t)$ 
around each adopted $\eta$ value.
For each synthetic spectrum we compute the instantaneous spectral slope
as a function of frequency and time:
$\alpha(\nu,t)\equiv \partial \log L_{\nu}/\partial \log \nu$.
We characterize the continuum SED variability by the
$\beta(\nu)\equiv {\partial \alpha(\nu)}/{\partial \log L_{\nu}}$  
parameter representing the variation of the local spectral slope per 
unit change of $\log L_{\nu}$.

Variability of QSO spectra has been analyzed on the basis of the
data of Giveon et al. (1999), consisting
of the light curves of a sample of 42
nearby and bright QSOs ($z<0.4, B< 16$ mag), belonging to the Palomar-Green 
(PG) sample,  monitored for 7 years
with median sampling interval of 39 days, in 
$B$ and $R$, with the 1 m telescope of the Wise Observatory.
For each time interval $\tau$ between two observations, it is possible 
to define
$\beta(\tau)\equiv {[\alpha(t+\tau)-\alpha(t)]}/{
[\log L_{\overline{\nu}}(t+\tau)-\log L_{\overline{\nu}}(t)]}$,
$\overline{\nu}= \sqrt{\nu_B \nu_R}$.
We consider, for each object,
the spectral variability parameter $\beta_m$ defined as the average of
$\beta(\tau)$ values for all $\tau \leq 1000$ days and
the average value of the spectral slope $\alpha_m$.
Both these  quantities can be compared with the values ($\beta$,$\alpha$)
obtained from the synthetic spectrum.
In Figure 1 the continuous curve represents $\beta$ versus $\alpha$ 
for  $\eta$ ranging from -3 (dominant host galaxy, in $B$ and $R$)
to 3 ($\sim$ pure QSO). To evaluate the effect of redshift we also show 
the dashed curve computed for $z=0.4$, the maximum redshift of the
Giveon et al. (1999) sample. 
On average, the distribution of points is clearly shifted towards top right
of Figure 1. This means that the relative weight of nuclear and stellar
component cannot account for the observed spectral variability, which
therefore must be intrinsic of the active nucleus.

%%%%%% fig 1 
\begin{figure}[ht]
\vspace*{-0.5cm}
\epsfxsize=12cm % fix the x-dimension and scales y-dim. to x-dim.
% Feel free to do the choice you prefer but do not exceed the
% x-dimension of the text lines. for centering: act on hspace argument.
%~\hspace{-0.5cm}
\epsfbox{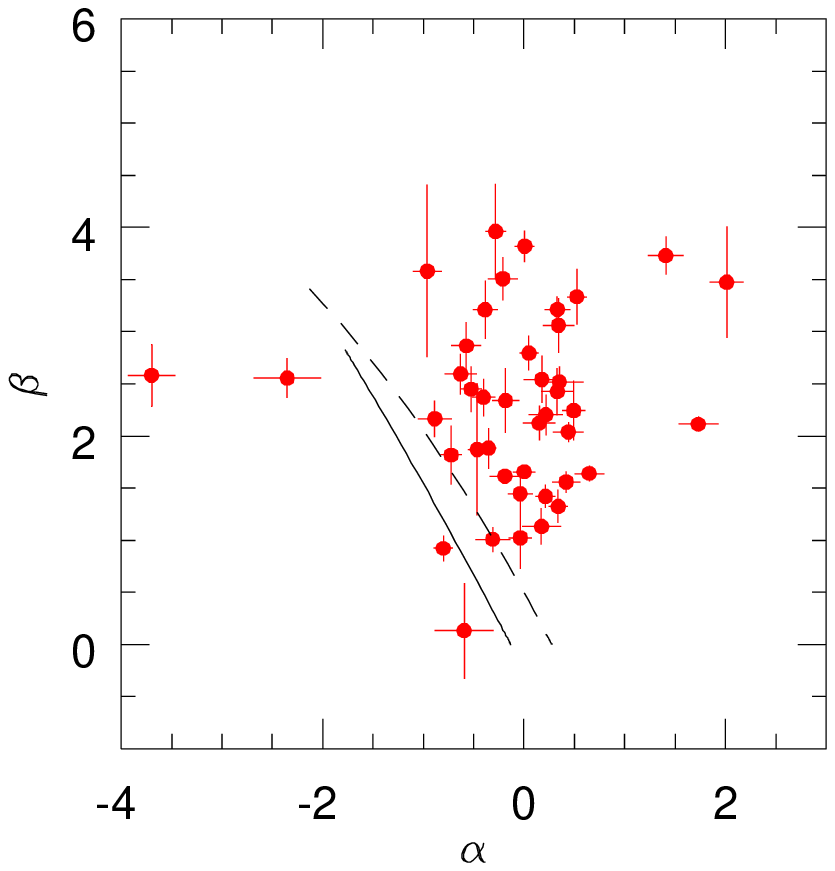}
\vspace*{-10.cm}

{\narrowcaption{
The observed $\beta$ versus $\alpha$ values
for PG QSOs. The two curves, ($z=0$ continuous, z=0.4 dashed),
represent $\beta(\alpha)$ of the composite spectrum, for $-3 < \eta < +3$ .}}
\end{figure}

\smallskip
We have computed synthetic QSO+host SEDs on the basis of templates of the
nuclear and the star light spectra. 
We have computed the spectral slope 
$\alpha$ and the spectral variability $\beta$
for the sample of PG QSOs monitored in the 
$B$ and $R$ bands by the Wise Observatory group. 
A comparison of the observed distribution of points in the $\beta-\alpha$
plane with the synthetic $\beta(\alpha)$ allows to disprove the hypothesis
that the observed variations of the QSO spectral shape are solely due
to a change of the relative weights of the nuclear and stellar component.
In the framework of the  analysis of Cid Fernandes et al. (2000), our result
implies that the constant spectral component
should be identified with a non flaring part of the accretion disk.

\end{document}